\documentclass{article}
\usepackage{spconf,amsmath,graphicx}
\usepackage{amsfonts}
\usepackage[table]{xcolor}
\usepackage{colortbl}
\usepackage{url}
\usepackage{multirow}
\usepackage{booktabs}
\usepackage{xcolor}
\usepackage{makecell}  

\title{AlignVSR: Audio-Visual Cross-modal Alignment for Visual Speech Recognition}
\name{Zehua Liu$^{1}$, Xiaolou Li$^{1}$, Chen Chen$^{2}$, Li Guo$^{1*}$, Lantian Li$^{1*}$, Dong Wang$^{2}$
}
\address{$^1$School of Artificial Intelligence, Beijing University of Posts and Telecommunications, China \\
$^2$Center for Speech and Language Technologies, BNRist, Tsinghua University, China}
\begin{document}
%
\maketitle
\begin{abstract}

Visual Speech Recognition (VSR) aims to recognize corresponding text by analyzing visual information from lip movements. 
Due to the high variability and weak information of lip movements, VSR tasks require effectively utilizing any information from any source and at any level. 
In this paper, we propose a VSR method based on audio-visual cross-modal alignment, named AlignVSR.
The method leverages the audio modality as an auxiliary information source and utilizes the global and local correspondence between the audio and visual modalities to improve visual-to-text inference. 
Specifically, the method first captures global alignment between video and audio through a cross-modal attention mechanism from video frames to a bank of audio units. 
Then, based on the temporal correspondence between audio and video, a frame-level local alignment loss is introduced to refine the global alignment, improving the utility of the audio information.
Experimental results on the LRS2 and CNVSRC.Single datasets consistently show that AlignVSR outperforms several mainstream VSR methods, demonstrating its superior and robust performance.
\end{abstract}
\section{Introduction}
\label{sec:intro}

Visual Speech Recognition (VSR), commonly known as lip reading, is a technology that utilizes the movements of the lips to recognize speech content~\cite{assael2016lipnet,shillingford2018large,prajwal2022sub}. 
This technology has vast potential in various fields, including public safety, assistance for the elderly and disabled, and video tampering detection.
One of the key challenges in VSR lies in the high variability and weak information provided by lip movements, demonstrated by our daily life experience that understanding others by reading lips is very hard. This challenge is particularly pronounced when dealing with homophones where different words have similar pronunciations. 
As a result, it is crucial to leverage information from any available source and at multiple levels to improve the performance of VSR systems.

In recent years, considering the strong information intersection between the audio modality and the visual and text modalities, many researchers have explored using the audio modality as an auxiliary information source to bridge the information gap between visual and textual information, thereby enhancing visual-to-text inference. 
For example, some studies have employed pre-trained Automatic Speech Recognition (ASR) models to produce audio features and then distill knowledge from the audio modality to the visual modality. 
For instance, Ma et al.~\cite{ma2022visual} introduced an L1 loss in their Hybrid CTC/Attention-based VSR model~\cite{watanabe2017hybrid} to force the visual encoder to reconstruct audio features, achieving temporal feature alignment. 
Similarly, Djilali et al.~\cite{djilali2023lip2vec} and Laux et al.~\cite{laux2024litevsr} proposed a similar approach that aligns audio and video at the feature level, but unlike Ma et al., they replaced the visual feature decoder with a pre-trained audio feature decoder during decoding.

To eliminate semantic irrelevant variation in acoustic features, recent research has explored quantizing audio features through clustering before aligning them with visual features. 
Yeo et al.~\cite{yeo2024akvsr} proposed an AKVSR approach, which first extracted audio features using the pre-trained Hubert~\cite{hsu2021hubert} model, then applied K-means clustering to quantize the features into a bank of audio units. Cross-modal attention was then employed to align each video frame to the bank of audio units. 
Ahn et al.~\cite{ahn2024syncvsr} employed the temporal correspondence between video and audio, and let the video frames predict the corresponding audio units. Their work reported the state-of-the-art (SOTA) performance.

\begin{figure*}
    \centering
    \includegraphics[width=1\linewidth]{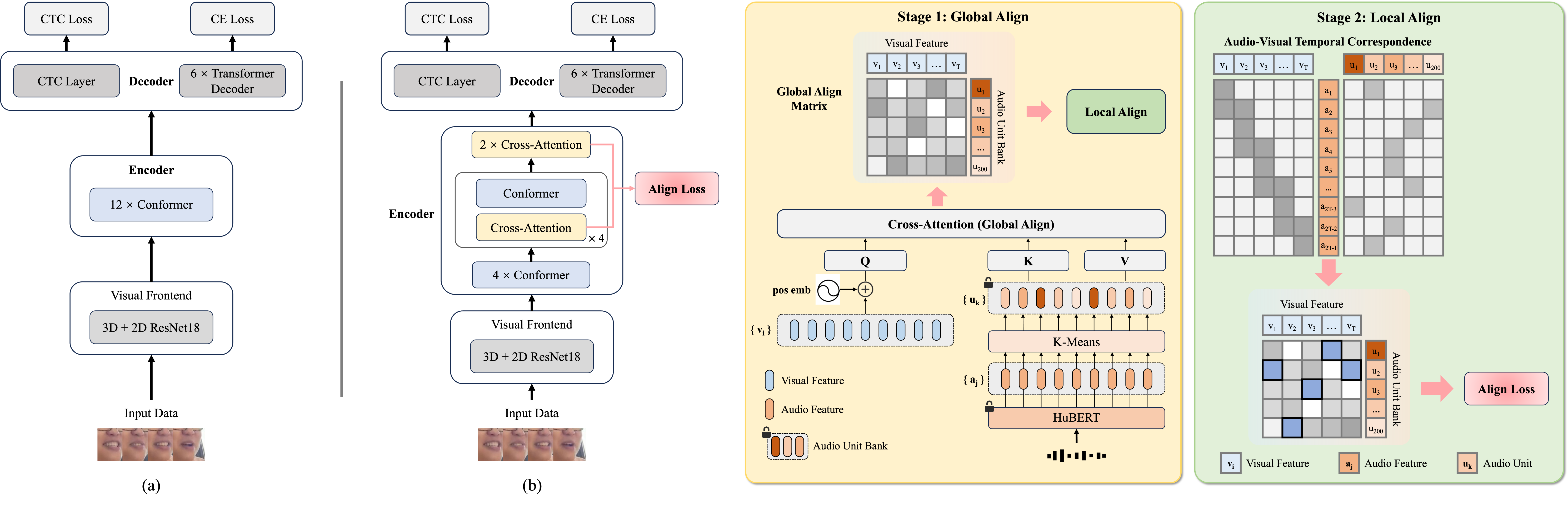}
    \caption{The framework of AlignVSR.}
    \label{fig:framework}
\end{figure*}

Inspired by the aforementioned approaches, this paper proposes a novel audio-visual cross-modal alignment method for VSR, named AlignVSR. 
The method follows the idea of distilling audio information, but designs a two-layer alignment mechanism to make the distillation effective. The first layer is a \emph{coarse global alignment} that aligns each video frame with a bank of audio units.
This is similar to AKVSR~\cite{yeo2024akvsr} in spirit, though we introduce a positional encoding to enhance contextual information. The second layer is a \emph{fine-grained local alignment} that aligns each video frame to the audio units of the corresponding audio frames. The idea of using audio-video temporal correspondence is similar to in~\cite{ahn2024syncvsr}, whereas our approach is built upon the global alignment architecture and aims to strengthen the attention weights on corresponding audio units rather than a strict local alignment. 
Extensive experiments on the LRS2~\cite{son2017lip} and CNVSRC.Single~\cite{chen2024cnvsrc} datasets demonstrate that our proposed AlignVSR consistently outperforms the AKVSR benchmark and several mainstream methods, achieving considerable performance improvements.

\section{Our Method: AlignVSR}

\subsection{Baseline System}
\label{sec:baseline}

Before introducing the proposed AlignVSR method, we briefly outline the baseline system. 
The baseline system adopts a mainstream end-to-end VSR model architecture~\cite{ma2021end,ma2022visual}, as shown in Fig.\ref{fig:framework}(a). 
The visual frontend uses the ResNet18 backbone~\cite{stafylakis2017combining}, while the encoder consists of a 12-layer Conformer network~\cite{gulati2020conformer} for extracting visual features. The decoder includes a Connectionist Temporal Classification (CTC) layer~\cite{graves2006connectionist} and 6 Transformer layers, with the Hybrid CTC-Attention training objective.

\subsection{AlignVSR}

Inspired by the success of the AKVSR approach~\cite{yeo2024akvsr}, the main idea of AlignVSR is to leverage the audio modality as an auxiliary information source to improve visual-to-text inference. 
The architecture contains a coarse global alignment like~\cite{yeo2024akvsr} and a fine-grained local alignment, as shown in Fig.\ref{fig:framework}(b) and detailed below.

\subsubsection{Global Alignment}

This stage closely follows the structure and process in~\cite{yeo2024akvsr}. 
Specifically, we first extract audio features using a pre-trained Hubert model and then apply K-means clustering to quantize the features into trainable audio units. 
Consistent with AKVSR, we set the number of clusters to 200. We then train the audio unit bank (comprising 200 audio units) using a Conformer encoder with Hybrid CTC/Attention loss in an ASR training framework.
The alignment is implemented as a cross-attention process from each video frame to the bank of audio units, i.e., the video features are query (Q), while the audio units serve as the key (K) and value (V). This alignment is `global' as a video frame can be aligned to any audio unit. 

To better preserve the video's temporal information, we involve the positional encoding in the video features (see ``pos emb" in stage 1 of Fig.\ref{fig:framework}(b)). This is different from AKVSR, and we empirically found performance gains from this new design.

\subsubsection{Local Alignment}

The global alignment does not exploit the temporal alignment between the audio and video modalities that we argue is an important information source. We, therefore, design a local alignment mechanism to tackle this problem, as shown in stage 2 of Fig.\ref{fig:framework}(b).


To make it clear, let us define the audio unit bank as $U=\{u_1, u_2,..., u_K\}$, where $K=200$ represents the number of quantized audio units. 
Let $V=\{v_1, v_2, ..., v_T\}$ denote the video feature sequence, where $T$ is the total number of video frames and $v_i$ represents the $i$-th video frame.
In this work, since the sampling rates of video and audio frames are different, with one video frame corresponding to three audio frames, the audio feature sequence is denoted as $A=\{a_1, a_2,..., a_{2T-1}\}$.
For each video frame $v_i$, the global attention scores on the audio units are defined by $P_i=\{p_i^{u_1}, p_i^{u_2}, ..., p_i^{u_K}\}$, 
where $p_i^{u_k}$ represents the attention score of the $i$-th video frame on the $k$-th audio unit $u_k$.

For video frame $v_i$, its corresponding audio frames are \{$a_{2i-2}$, $a_{2i-1}$, $a_{2i}$\}, and the associated audio units are denoted by \{$u_f$, $u_s$, $u_t$\}. We define the Align loss as follows
\begin{equation}
\mathcal{L}_i = - \log p_{i}^{u_f} - \log p_{i}^{u_s} - \log p_{i}^{u_t},
\end{equation}
\noindent which is essentially a cross entropy between $P_i$ and an indicator vector $I\in \{0,1\}^{200}$ for while the elements at $u_f, u_s, u_t$ are 1 and all other elements are 0.

The final Align loss is the average loss over all video frames, formulated as:
\begin{equation}
\mathcal{L}_{\text{Align}} = \frac{1}{T}\sum_{i=1}^{T} L_i
\end{equation}

\subsubsection{Loss Function}

The final loss includes the Hybrid CTC-Attention loss and the Align loss, with appropriate weights set empirically:

\begin{equation}
\mathcal{L} = \alpha \mathcal{L}_{\text{CTC}} + \beta \mathcal{L}_{\text{Att}} + \gamma \mathcal{L}_{\text{Align}}
\end{equation}
\noindent where $L_{\text{CTC}}$ is the CTC loss, $L_{\text{Att}}$ is the Attention loss, and $L_{\text{Align}}$ is our proposed Align loss. 
$\alpha$, $\beta$, and $\gamma$ are hyperparameters that control the relative importance of each loss component in the final objective.

\section{Experimental Setup}

\subsection{Datasets}
To evaluate the performance of AlignVSR, we conducted experiments on two datasets: the English LRS2 dataset~\cite{son2017lip} and the Chinese CNVSRC.Single dataset~\cite{chen2024cnvsrc}. 
The LRS2 dataset consists of 144,482 video clips, totaling 225 hours of video, and is split into three subsets: training (195 hours), validation (29 hours), and testing (0.5 hours). 
The CNVSRC.Single dataset, released for the CNVSRC 2023\footnote{https://cnceleb.org/competition}, is a single-speaker dataset containing a training set of 83 hours and a test set of 10 hours.

\subsection{Data Preprocessing}
Our data preprocessing pipeline follows AUTO-AVSR~\cite{ma2023auto}. 
First, we employ RetinaFace~\cite{deng2019retinaface} to detect the face region in each video frame. 
Next, facial landmarks are extracted using the Facial Alignment Network (FAN)~\cite{bulat2017far}, and these landmarks are used to align the detected faces to a canonical reference face. 
Finally, the lip region is cropped from each frame to a resolution of 96×96 pixels and used as input to the model.

\subsection{Model Architecture}
First, the visual frontend follows the ResNet18 backbone in~\cite{stafylakis2017combining}. 
Specifically, the input video passes through a 3D-CNN layer and 8 2D-CNN layers to capture local spatio-temporal features. 
Then, the encoder adopts the Conformer architecture~\cite{gulati2020conformer}, and we involve the cross-attention module in aligning each visual feature frame with the audio unit bank, producing the final visual features. 
Finally, the decoder consists of a Connectionist Temporal Classification (CTC) layer and 6 Transformer layers to decode the visual features into text. Fig.1(b) illustrates the detailed model architecture.

\subsection{Training Recipe}
We employed a curriculum learning approach for model training, drawing on the training strategies used in the CNVSRC 2023~\cite{chen2024cnvsrc}. 
The training processes on LRS2 and CNVSRC.Single datasets are described as follows:

\begin{itemize}
\item \textbf{LRS2}: We first extracted the video clips from the training set, which are shorter than 4 seconds, and trained an initial model. 
Then, the model was fine-tuned on the full training set to obtain the final model.
\item \textbf{CNVSRC.Single}: We first split the training set into video clips shorter than 4 seconds for initial training. After this initial phase, the trained model was used as the pre-trained model for further training on the full training set.
\end{itemize}

In our experiments, the values of $\alpha$, $\beta$, and $\gamma$ are empirically set to 0.08, 0.9, and 6.5 in the LRS2 dataset and 0.085, 0.9, and 3.5 in the CNVSRC.Single dataset. The source code is available at~\url{https://github.com/liu12366262626/AlignVSR}.

\section{Results}

\subsection{Basic Results}

First, we validate the effectiveness of the AlignVSR model on both the LRS2 and CNVSRC.Single datasets. 
For a quick validation, we extracted video clips shorter than 5 seconds from the LRS2.Train dataset (a total of 69 hours) to construct an LRS2.Train subset for model training.

We constructed three VSR systems for comparison:
\begin{itemize}
\item \textbf{S1}: A baseline system based on the Conformer architecture, as described in Section~\ref{sec:baseline}.
\item \textbf{S2}: A system following the AKVSR approach~\cite{yeo2024akvsr}, where a cross-attention module is involved to the baseline Conformer system to achieve global alignment between audio and video modalities. Notably, unlike standard AKVSR, we incorporate positional encoding into the original video features to better capture their temporal correlation.
\item \textbf{S3}: Building upon S2, this system introduces our proposed Align loss during training to achieve global and local cross-modal alignment between audio and video modalities.
\end{itemize}

The experimental results are shown in Table~\ref{tab:basic} below. First, comparing S1 and S2, we observe that S2 significantly outperforms S1 on both datasets. This indicates that incorporating auxiliary information from the audio modality enhances visual-to-text inference, a finding consistent with AKVSR's conclusions. 

Next, comparing S2 and S3, we see that S3 achieves overwhelming improvements, particularly on the LRS2 test set. This demonstrates that global cross-modal alignment alone cannot fully align audio and video modalities. By incorporating the proposed global and local alignment mechanism, S3 more effectively leverages the temporal correspondence between audio and video, resulting in superior alignment performance.

\begin{table}[h]
\centering
\caption{Basic results with three systems.}
\label{tab:basic}
\resizebox{1.0\columnwidth}{!}{
\begin{tabular}{l|l|c|c}
\toprule
\multirow{2}{*}{System} & \multirow{2}{*}{Method} & LRS2.Test & CNVSRC.Single.Test \\
  &   &  WER(\%)   & CER(\%)  \\
\midrule
S1  &  Conformer (Baseline) & 66.75 & 49.92 \\ 
S2 &  Conformer + Cross-Att & 64.30 & 48.12 \\ 
\midrule
S3 &  S2 + Align Loss (Ours) & \textbf{45.63} & \textbf{46.06} \\ 
\bottomrule
\end{tabular}}
\end{table}

To better illustrate the effect of the proposed Align loss, we plot the changes in CTC loss, Attention loss, and Align loss for both S2 and S3 during the training process, as shown in Fig.\ref{fig:loss} below.
It is evident that for S2, since no explicit audio-visual alignment information is incorporated, the Align loss remains relatively stable during the training process. In contrast, for S3, the Align loss shows a consistent downward trend, indicating that the Align loss effectively facilitates the alignment between audio and video modalities. This observation suggests that relying solely on implicit global cross-attention is insufficient for effective audio-visual alignment. 

More interestingly, the inclusion of Align loss also accelerates the reduction of both the CTC loss and the Attention loss. 
This demonstrates that the proposed Align loss indeed facilitates the alignment between audio and video modalities, thereby further enhancing visual-to-text inference.

\begin{figure}[h]
    \centering
    \includegraphics[width=1\linewidth]{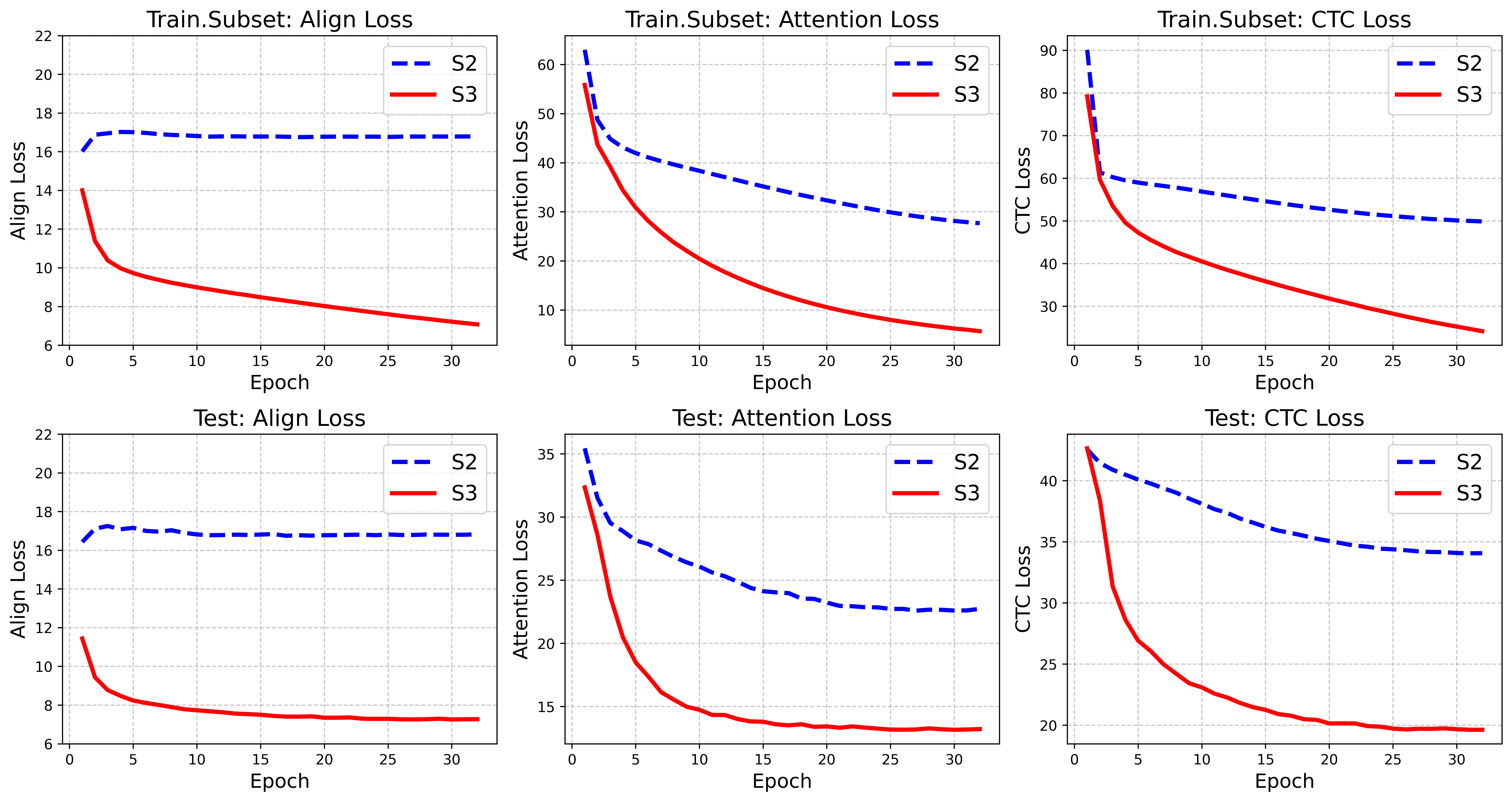}
    \caption{The change of three losses on both the training set and test set of LRS2 during the model training.}
    \label{fig:loss}
\end{figure}

\subsection{Comparison with other systems}

Next, we compare our method with two mainstream audio modality auxiliary approaches. The first is by Ma et al.~\cite{ma2022visual}, who used a pre-trained ASR model and employed an L1 loss function to achieve temporal alignment between audio and video at the feature level. 
Due to the significant differences between audio and video feature spaces, directly aligning them at the feature level poses challenges. 
To address this, Ahn et al.~\cite{ahn2024syncvsr} proposed using vector quantization and align video and audio features by predicting audio units from the corresponding video features. This is also used as a competitive method. 

To ensure a fair comparison, we trained the AlignVSR system using the same LRS2 training set as \cite{ma2022visual} and \cite{ahn2024syncvsr}, and evaluated it on the same LRS2 test set. The experimental results are presented in Table~\ref{tab:power} below.

\begin{table}[h]
\centering
\caption{Comparison with more powerful systems on LRS2.}
\label{tab:power}
\begin{tabular}{l|c|c}
\toprule
Method                    & Labeled data  & WER (\%) \\ 
\midrule
Ma~\cite{ma2022visual}    & 223h          & 32.9 \\ 
Ahn~\cite{ahn2024syncvsr} & 223h          & 30.7 \\ 
\midrule
\textbf{Ours}             & 223h          & \textbf{30.2} \\
\bottomrule
\end{tabular}
\end{table}

It can be seen that our proposed method achieves the best performance. Notably, compared to Ahn~\cite{ahn2024syncvsr}, which also utilizes audio-video correspondence, our method demonstrates some advantages. Our conjecture is that Ahn~\cite{ahn2024syncvsr} performs strict frame-level local alignment, whereas our method builds on global alignment, and the local correspondence knowledge is only used to refine the alignment. We believe this soft alignment is more reasonable as the audio and video streams are not strictly aligned in time.

\section{Conclusion}

In this paper, we propose a VSR method based on fine-grained audio-visual cross-modal alignment, named AlignVSR. The method uses the temporal correspondence between audio and video to regulate the cross-attention between video features and a bank of audio units in the form of an extra alignment loss. 
Experiments on the LRS2 and CNVSRC.Single datasets demonstrate that our proposed AlignVSR consistently outperforms the AKVSR benchmark and several other mainstream methods, achieving significant performance improvements. Moreover, the new method substantially improves model convergence and thus eases the training. In the future, we plan to validate our method on more datasets and integrate diverse information through the soft alignment structure. 

\bibliographystyle{IEEEbib}
\bibliography{refs}

\begin{thebibliography}{10}

\bibitem{assael2016lipnet}
Yannis~M Assael, Brendan Shillingford, Shimon Whiteson, and Nando De~Freitas,
\newblock ``Lipnet: End-to-end sentence-level lipreading,''
\newblock {\em arXiv preprint arXiv:1611.01599}, 2016.

\bibitem{shillingford2018large}
Brendan Shillingford, Yannis Assael, Matthew~W Hoffman, Thomas Paine, C{\'\i}an
  Hughes, Utsav Prabhu, Hank Liao, Hasim Sak, Kanishka Rao, Lorrayne Bennett,
  et~al.,
\newblock ``Large-scale visual speech recognition,''
\newblock {\em arXiv preprint arXiv:1807.05162}, 2018.

\bibitem{prajwal2022sub}
KR~Prajwal, Triantafyllos Afouras, and Andrew Zisserman,
\newblock ``Sub-word level lip reading with visual attention,''
\newblock in {\em Proceedings of the IEEE/CVF conference on Computer Vision and
  Pattern Recognition}, 2022, pp. 5162--5172.

\bibitem{ma2022visual}
Pingchuan Ma, Stavros Petridis, and Maja Pantic,
\newblock ``Visual speech recognition for multiple languages in the wild,''
\newblock {\em Nature Machine Intelligence}, vol. 4, no. 11, pp. 930--939,
  2022.

\bibitem{watanabe2017hybrid}
Shinji Watanabe, Takaaki Hori, Suyoun Kim, John~R Hershey, and Tomoki Hayashi,
\newblock ``Hybrid ctc/attention architecture for end-to-end speech
  recognition,''
\newblock {\em IEEE Journal of Selected Topics in Signal Processing}, vol. 11,
  no. 8, pp. 1240--1253, 2017.

\bibitem{djilali2023lip2vec}
Yasser Abdelaziz~Dahou Djilali, Sanath Narayan, Haithem Boussaid, Ebtessam
  Almazrouei, and Merouane Debbah,
\newblock ``Lip2vec: Efficient and robust visual speech recognition via
  latent-to-latent visual to audio representation mapping,''
\newblock in {\em Proceedings of the IEEE/CVF International Conference on
  Computer Vision}, 2023, pp. 13790--13801.

\bibitem{laux2024litevsr}
Hendrik Laux, Emil Mededovic, Ahmed Hallawa, Lukas Martin, Arne Peine, and Anke
  Schmeink,
\newblock ``Litevsr: Efficient visual speech recognition by learning from
  speech representations of unlabeled data,''
\newblock in {\em ICASSP 2024-2024 IEEE International Conference on Acoustics,
  Speech and Signal Processing (ICASSP)}. IEEE, 2024, pp. 10391--10395.

\bibitem{yeo2024akvsr}
Jeong~Hun Yeo, Minsu Kim, Jeongsoo Choi, Dae~Hoe Kim, and Yong~Man Ro,
\newblock ``Akvsr: Audio knowledge empowered visual speech recognition by
  compressing audio knowledge of a pretrained model,''
\newblock {\em IEEE Transactions on Multimedia}, 2024.

\bibitem{hsu2021hubert}
Wei-Ning Hsu, Benjamin Bolte, Yao-Hung~Hubert Tsai, Kushal Lakhotia, Ruslan
  Salakhutdinov, and Abdelrahman Mohamed,
\newblock ``Hubert: Self-supervised speech representation learning by masked
  prediction of hidden units,''
\newblock {\em IEEE/ACM transactions on audio, speech, and language
  processing}, vol. 29, pp. 3451--3460, 2021.

\bibitem{ahn2024syncvsr}
Young~Jin Ahn, Jungwoo Park, Sangha Park, Jonghyun Choi, and Kee-Eung Kim,
\newblock ``Syncvsr: Data-efficient visual speech recognition with end-to-end
  crossmodal audio token synchronization,''
\newblock {\em arXiv preprint arXiv:2406.12233}, 2024.

\bibitem{son2017lip}
Joon Son~Chung, Andrew Senior, Oriol Vinyals, and Andrew Zisserman,
\newblock ``Lip reading sentences in the wild,''
\newblock in {\em Proceedings of the IEEE conference on computer vision and
  pattern recognition}, 2017, pp. 6447--6456.

\bibitem{chen2024cnvsrc}
Chen Chen, Zehua Liu, Xiaolou Li, Lantian Li, and Dong Wang,
\newblock ``Cnvsrc 2023: The first chinese continuous visual speech recognition
  challenge,''
\newblock {\em arXiv preprint arXiv:2406.10313}, 2024.

\bibitem{ma2021end}
Pingchuan Ma, Stavros Petridis, and Maja Pantic,
\newblock ``End-to-end audio-visual speech recognition with conformers,''
\newblock in {\em ICASSP 2021-2021 IEEE International Conference on Acoustics,
  Speech and Signal Processing (ICASSP)}. IEEE, 2021, pp. 7613--7617.

\bibitem{stafylakis2017combining}
Themos Stafylakis and Georgios Tzimiropoulos,
\newblock ``Combining residual networks with lstms for lipreading,''
\newblock {\em arXiv preprint arXiv:1703.04105}, 2017.

\bibitem{gulati2020conformer}
Anmol Gulati, James Qin, Chung-Cheng Chiu, Niki Parmar, Yu~Zhang, Jiahui Yu,
  Wei Han, Shibo Wang, Zhengdong Zhang, Yonghui Wu, et~al.,
\newblock ``Conformer: Convolution-augmented transformer for speech
  recognition,''
\newblock {\em arXiv preprint arXiv:2005.08100}, 2020.

\bibitem{graves2006connectionist}
Alex Graves, Santiago Fern{\'a}ndez, Faustino Gomez, and J{\"u}rgen
  Schmidhuber,
\newblock ``Connectionist temporal classification: labelling unsegmented
  sequence data with recurrent neural networks,''
\newblock in {\em Proceedings of the 23rd international conference on Machine
  learning}, 2006, pp. 369--376.

\bibitem{ma2023auto}
Pingchuan Ma, Alexandros Haliassos, Adriana Fernandez-Lopez, Honglie Chen,
  Stavros Petridis, and Maja Pantic,
\newblock ``Auto-avsr: Audio-visual speech recognition with automatic labels,''
\newblock in {\em ICASSP 2023-2023 IEEE International Conference on Acoustics,
  Speech and Signal Processing (ICASSP)}. IEEE, 2023, pp. 1--5.

\bibitem{deng2019retinaface}
Jiankang Deng, Jia Guo, Yuxiang Zhou, Jinke Yu, Irene Kotsia, and Stefanos
  Zafeiriou,
\newblock ``Retinaface: Single-stage dense face localisation in the wild,''
\newblock {\em arXiv preprint arXiv:1905.00641}, 2019.

\bibitem{bulat2017far}
Adrian Bulat and Georgios Tzimiropoulos,
\newblock ``How far are we from solving the 2d \& 3d face alignment
  problem?(and a dataset of 230,000 3d facial landmarks),''
\newblock in {\em Proceedings of the IEEE international conference on computer
  vision}, 2017, pp. 1021--1030.

\end{thebibliography}

\end{document}